\documentclass[a4paper,10pt,twoside]{cpc-hepnp}
\usepackage{multicol}
\usepackage{graphicx}
\usepackage{booktabs}
\usepackage{amssymb,bm,mathrsfs,bbm,amscd}
\usepackage[tbtags]{amsmath}
\usepackage{lastpage}
\usepackage{CJK}
\usepackage{epstopdf}
\usepackage{epsfig}
\usepackage{color}
\usepackage{ifpdf}

\usepackage{ulem}

\newcommand{\delete}{\bgroup\markoverwith{\textcolor{blue}{\rule[0.5ex]{2pt}{1pt}}}\ULon}

\begin{document}

\fancyhead[c]{\small Chinese Physics C~~~Vol. $\times\times$, No. 1 ($\times\times\times\times$)
010201} \fancyfoot[C]{\small 010201-\thepage}
\footnotetext[0]{Received 14 March 2009}
\title{Restoration of pseudo-spin symmetry in $N=32$ and $34$ isotones described by relativistic Hartree-Fock theory\thanks{Supported by National Natural Science Foundation of China (11675065, 11711540016) }}
\author{%
      Zheng Zheng Li%
\quad Shi Yao Chang%
\quad Qiang Zhao
\quad Wen Hui Long$^{1)}$\email{longwh@lzu.edu.cn}%
\quad Yi Fei Niu
}
\maketitle
\address{%
 School of Nuclear Science and Technology, Lanzhou University, Lanzhou 730000, China\\
}

\begin{abstract}
Restoration of pseudo-spin symmetry (PSS) along the $N=32$ and $34$ isotonic chains and the physics behind are studied by applying the relativistic Hartree-Fock theory with effective Lagrangian PKA1. Taking the proton pseudo-spin partners $(\pi2s_{1/2},\pi1d_{3/2})$ as candidates, systematic restoration of PSS along both isotonic chains is found from sulphur (S) to nickel (Ni), while distinct violation from silicon (Si) to sulphur is discovered near the drip lines. The effects of the tensor-force components introduced naturally by the Fock terms are investigated, which can only partly interpret the systematics from calcium to nickel, but fail for the overall trends. Further analysis following the Schr\"{o}dinger-like equation of the lower component of Dirac spinor shows that the contributions from the Hartree terms dominate the overall systematics of the PSS restoration, and such effects can be self-consistently interpreted by the evolution of the proton central density profiles along both isotonic chains. Specifically the distinct PSS violation is found to tightly relate with the dramatic changes from the bubble-like density profiles in silicon to the central-bumped ones in sulphur.
\end{abstract}
\footnotetext[0]{\hspace*{-3mm}\raisebox{0.3ex}{$\scriptstyle\copyright$}2013
Chinese Physical Society and the Institute of High Energy Physics
of the Chinese Academy of Sciences and the Institute
of Modern Physics of the Chinese Academy of Sciences and IOP Publishing Ltd}%

\begin{multicols}{2}

\section{Introduction}

The pseudo-spin symmetry (PSS), firstly found in spherical nuclei \cite{Arima-1969-PLB, Hechet-1969-NPA} and later in deformed ones \cite{A.Bohr-1982-P.S., Beuschel-1997-NPA}, corresponds to the quasi-degeneracy of two single-particle orbits $(n,$ $l$, $j=l+1/2)$ and ($ n -1$, $ l+2$, $  j=l+3/2$), namely the pseudo-spin partners $j_\gtrless =\tilde l \pm 1/2$ with $\tilde l = l+1$. Since the inchoation of PSS, intensive efforts have been devoted to exploring the origin and conservation condition of PSS in the past decades \cite{Ginocchio-1997-PRL, Ginocchio-1999-P.R., Ginocchio-2005-P.R., H.Z.Liang-2015-P.R., J.Meng-1998-PRC, W.H.Long-2006-PLB-PSS, Sugawara-1998-PRC}. Until the end of last century, the PSS is recognized as a relativistic symmetry \cite{Ginocchio-1997-PRL}. The conservation condition of PSS is that the sum of the scalar $S(r)$ and vector $V(r)$ potentials in the single-particle Dirac equation equals to zero, i.e., $V(r)+S(r) = 0$ \cite{Ginocchio-1997-PRL}. Later, it was also demonstrated that the exact PSS can be deduced with the condition $d(V+S)/dr =0$ \cite{J.Meng-1998-PRC}, and further, a new relativistic model that fulfills such condition with bound single-particle states was raised up in Ref. \cite{T.S.Chen-2003-CPL}. Moreover, it becomes well known that the orbital angular momentum $\tilde l$ is nothing but the one of the lower components of the Dirac spinors \cite{Ginocchio-1997-PRL}. Afterwards, the PSS has been also tested in such as deformed nuclei \cite{Ginocchio-2004-PRC} and resonant states \cite{B.N.Lv-2012-PRL, N.Li-2016-PRL} within the relativistic frame. Besides, the conservation condition also leads to rather precisely preserved spin symmetry (SS) in anti-nucleon spectra, which indeed shares the origin with the PSS in nuclear single particle spectra \cite{S.G.Zhou-2003-PRL}. As revealed by the conservation conditions, both the PSS and SS indeed reflect an important nature of nuclear dynamics, which benchmarks our understanding on the nuclear structure properties.

Over the past few decades, worldwide rapid development of the radioactive ion beam (RIB) facilities and detector systems has brought about the revolution of nuclear physics and largely enriched our knowledge of atomic nucleus from the stable ones to the ones far from the stability valley. When approaching the neutron/proton drip lines, unexpected characteristic structures such as new magic numbers can arise \cite{Sorlin-2008-PPNP}, for instance, the neutron magicity $N=16$ observed in drip-line doubly magic nucleus $^{24}$O \cite{Hoffman-2008-PRL}. More recently the new magicities at $N=32$ and $34$ in calcium isotopes have been indicated from the experimental evidences, such as rather high excitation energy of the $2_1^+$ state and high precision mass measurements \cite{Dinca-2005-PRC, Gade-2006-PRC, Liddick-2004-PRL, Prisciandaro-2001-PLB, Wienholtz-2013-Nature, Steppenbeck-2013-Nature}. The open of new magicities at $N=32$ and $34$ was further supported by the ab-initio calculations \cite{Hagen-2012-PRL, Hergert-2014-PRC}, shell model \cite{Steppenbeck-2013-Nature, Steppenbeck-2015-PRL}, and some energy density functionals with the tensor force \cite{J.J.Li-2016-PLB, Grasso-2014-PRC, Yuksel-2014-PRC}. Specifically it was suggested that the nuclear tensor force may play a role in the occurrence of $N=32$ magicity in $^{52}$Ca \cite{J.J.Li-2016-PLB}. Very recently it was realized that the opening of $N=34$ magicity might be related to the emergence of the bubble-like structure in the predicted doubly magic drip line nucleus $^{48}$Si \cite{J.J.Li-2019-PLB} that in general leads to evident PSS violation \cite{Ginocchio-2005-P.R.,J.Meng-2006-PPNP,J.Meng-1998-PRC,J.Meng-1999-PRC}. It in fact inspires our interests to explore the restoration of PSS in the relevant nuclei, particularly the $N=32$ and $34$ isotones.

Relativistically the description of the nuclear dynamics of ground state can be achieved with the meson exchange diagram of nuclear force \cite{Yukawa-1935-Proc..}, following the spirit of density functional theory \cite{Hohenberg-1964-PRB, Kohn-1965-PRA}. It leads to the well-known covariant density functional (CDF) theory, including the main branches the relativistic mean field (RMF) \cite{Walecka-1974-Ann.Phys., Ring-1996-PPNP, J.Meng-2006-PPNP, J.Meng-1998-NPA, Reinhard-1989-PPNP} and relativistic Hartree-Fock (RHF) \cite{Bouyssy-1987-PRC, W.H.Long-2006-PLB-RHF, Bernardos-1993-PRC, Marcos-2004-JPG, B.Y.Sun-2008-PRC, W.H.Long-2012-PRC, W.H.Long-2007-PRC, W.H.Long-2010-PRC, J.J.Li-2016-PLB, J.J.Li-2019-PLB} theories, which have provided successful description of nuclear structure properties for nuclei in the whole nuclear chart. Aiming at the $N=32$ and $34$ isotones, we restrict ourselves within the RHF theory due to the facts that the presence of Fock terms has brought significant improvements in the self-consistent description of shell evolution \cite{W.H.Long-2008-EPL, W.H.Long-2009-PLB, L.J.Wang-2013-PRC, W.H.Long-2010-PRC}, nuclear tensor force \cite{W.H.Long-2008-EPL, L.J.Jiang-2015-PRC, Y.Y.Zong-2018-CPC, Z.H.Wang-2018-PRC, L.J.Wang-2013-PRC}, appropriate restoration of PSS \cite{W.H.Long-2006-PLB-PSS, W.H.Long-2007-PRC, H.Z.Liang-2010-EPJA, W.H.Long-2010-PRC, J.J.Li-2016-PLB}, nuclear spin-isospin excitations \cite{H.Z.Liang-2008-PRL, Z.M.Niu-2017-PRC}, $\beta$-decay properties \cite{Z.M.Niu-2008-PLB}, etc. Particularly, with the RHF Lagrangian PKA1 \cite{W.H.Long-2007-PRC}, in which the $\pi$- and $\rho$-tensor couplings have been taken into account, the new magicities $N=32$ and $34$ in $^{52, 54}$Ca can be well reproduced, being consistent with the experimental evidences and other theoretical calculations. It is worthwhile to mention that so far the RHF-PKA1 is the only covariant density functional that can interpret self-consistently the occurrence of both new magicities $N=32$ and $34$.

In present work, we will focus on the restoration of PSS along the isotonic chains of $N=32$ and $34$ and the physics behind, by adopting the RHF calculations with the RHF Lagrangian PKA1. The contents are organized as follows. In Sec. $2$, the general formalism of radial Dirac equation and the corresponding Schr\"{o}dinger-like one is given. In Sec. $3$, the restoration of PSS along the $N=32$ and $34$ isotonic chain are studied by taking the proton pseudo-spin doublet ($\pi2s_{1/2}$, $\pi1d_{3/2}$) as an example, in which the physics behind will be discussed in details. In the end, a brief summary is given in Sec. $4$.

\section{Radial Dirac equation and the deduced Schr\"{o}dinger-like one}

For the $N=32$ and $34$ isotones, the spherical symmetry is assumed in this study. One has to admit that some of the selected isotones are potentially deformed, however, the restriction of spherical symmetry is just for simplicity and the convenience in revealing the physics of the PSS violation and restoration. With the assumption of spherical symmetry, as we will see, similar systematics of the violation and restoration of the PSS are found for both $N=32$ and $34$ isotones, and following the similar density evolutions the mechanism is clarified.

For a spherical nucleus, the Dirac spinor of nucleon is of the following form as,
\begin{align}
\psi (\pmb r) = & \dfrac{1}{r} \begin{pmatrix} i G (r) \mathcal Y_{j m }^{l }(\vartheta,\varphi) \\[0.5em] -F (r) \mathcal Y_{j m }^{l '}(\vartheta,\varphi)  \end{pmatrix},
\end{align}
where $G(r)$ and $F(r)$ denote the radial wave functions for the upper and lower components, respectively, and $\mathcal Y_{jm}^l$ is the spherical spinor with angular momentum $j$ and the projection $m$. For the orbital angular momenta $l$ and $l'$ of the upper and lower components, one can have $l+l'=2j$. As mentioned, the pseudo-spin partners share the orbital angular momentum $l'$ of the lower component.

In the RHF approach, restricted to the spherical symmetry, the variation of the RHF energy functional \cite{W.H.Long-2006-thesis} leads to the radial Dirac equations, i.e. the relativistic Hartree-Fock equations as,
\begin{subequations}\label{Dir}
\begin{align}
E G(r) = & - \Big[\dfrac{d}{dr} - \dfrac{\kappa}{r} - \Sigma_T(r)\Big] F(r)\nonumber\\ & \hspace{1em} + \big[\Sigma_0(r) + \Sigma_S(r)\big] G(r) + Y(r),\\
E F(r) = & + \Big[\dfrac{d}{dr} + \dfrac{\kappa}{r} + \Sigma_T(r)\Big] G(r)\nonumber\\ & \hspace{1em} + \big[\Sigma_0(r) - \Sigma_S(r)-2M\big] F(r) + X(r),
\end{align}
\end{subequations}
where $E$ is the single-particle energy, $\kappa = j+1/2$ for $j=l-1/2$ and $-(j+1/2)$ for $j=l+1/2$, and $\Sigma_0$, $\Sigma_S$ and $\Sigma_T$ are the vector, scalar and tensor self-energies contributed by the Hartree and rearrangement terms \cite{W.H.Long-2006-thesis, Bouyssy-1987-PRC, W.H.Long-2006-PLB-RHF, W.H.Long-2007-PRC, W.H.Long-2010-PRC}. Different from the RMF theory, the inclusion of Fock terms leads to the non-local integral terms $Y$ and $X$ \cite{Bouyssy-1987-PRC, W.H.Long-2006-thesis}, and the radial Dirac equation become an integro-differential one. In order to avoid the complexity in solving integro-differential equation, the non-local integral term $Y$ is localized as
\begin{align}
  Y_G(r) \equiv & \dfrac{Y(r) G(r)}{G^2(r) + F^2(r)}, & Y_F(r) \equiv & \dfrac{Y(r) F(r)}{G^2(r) + F^2(r)},
\end{align}
and so does the term $X$ with equivalent local potentials $X_G$ and $X_F$ \cite{Bouyssy-1987-PRC}. Thus, the radial Dirac equation (\ref{Dir}) can be expressed as,
\begin{subequations}\label{Dir2}
\begin{align}
&\Big[\dfrac{d}{dr} - \dfrac{\kappa}{r} - \Sigma_T - Y_F\Big]F - \big[\Delta - E\big]G = 0,\\[0.5em]
&\Big[\dfrac{d}{dr} + \dfrac{\kappa}{r} + \Sigma_T + X_G\Big]G + \big[V - E\big]F=0,
\end{align}
\end{subequations}
where $\Delta \equiv  \Delta^D + Y_G$ with $\Delta^D \equiv  \Sigma_S + \Sigma_0$, and $V \equiv  V^D + X_F$ with $V^D\equiv  \Sigma_0 - \Sigma_S - 2M$.

Since the PSS is tightly related to the lower component of Dirac spinor, it is convenient in general to dervie a Schr\"{o}dinger-like equation for the $F$ component \cite{J.Meng-1998-PRC,W.H.Long-2006-PLB-PSS}, which can be expressed as,
\begin{align}
&\dfrac{1}{V^D - E} \dfrac{d^2}{dr^2} F + \dfrac{1}{V^D-E} \big[V_{PCB} + \hat{\mathcal{V}}^D + \hat{\mathcal{V}}^E \big] F = E F,\label{Sch-L}
\end{align}
where the pseudo centrifugal barrier $V_{PCB}$, and the Hartree ($\hat{\mathcal V}^D$) and Fock ($\hat{\mathcal V} ^E$) terms read as,
\begin{align}
&V _{PCB} = - \dfrac{\kappa(\kappa - 1)}{r ^2},\\
&\hat{\mathcal{V}}^D = V_1^D \dfrac{d}{dr} + V_2^D + V^D_{\text{PSO}} + V_\Delta,\\
&\hat{\mathcal{V}}^E = V_1^E \dfrac{d}{dr} + V_2^E + V^E_{\text{PSO}}.
\end{align}
In the Hartree ($\hat{\mathcal V}^D$) and Fock ($\hat{\mathcal V}^E$) terms, the relevant terms read as,
\begin{align}
&V_1^D = -\dfrac{1}{\Delta-E} \dfrac{d\Delta^D}{dr}, \\
&V_2^D = \dfrac{ \Sigma_T }{\Delta-E} \dfrac{d\Delta^D}{dr} - \dfrac{d\Sigma_T}{dr} - \Sigma_T^2,\\
&V_{\text{PSO}}^D = \dfrac{\kappa}{r} \Big[\dfrac{1}{\Delta-E} \dfrac{d\Delta^D}{dr}- 2 \Sigma_T \Big],\\
&V_{\Delta} = \Delta ^D (V ^D - E),\\
&V_1^E = -\dfrac{1}{\Delta-E} \dfrac{dY_G}{dr} + X_G - Y_F,\\
&V_2^E = \dfrac{1}{\Delta-E} \dfrac{dY_G}{dr} \Sigma_T + \dfrac{1}{\Delta-E} \dfrac{d\big(\Delta^D+Y_G\big)}{dr} Y_F \nonumber \\
&\qquad~  - \dfrac{dY_F}{dr} - \Sigma_T \big[X_G+Y_F\big] - X_G Y_F\\
&\qquad~  + (V^D - E) Y_G + X_F (\Delta - E), \nonumber\\
&V_{\text{PSO}}^E = \dfrac{\kappa}{r} \Big[\dfrac{1}{\Delta-E} \dfrac{dY_G}{dr} - X_G - Y_F\Big],
\end{align}
where $V_{\text{PSO}}$ corresponds to the pseudo-spin orbital (PSO) potential. Similar as Ref. \cite{W.H.Long-2006-PLB-PSS}, the following integral is introduced to evaluate the contributions from various channels to the single-particle energy $E$,
\begin{align}
\dfrac{1}{\int_0^\infty F^2 dr} \int_0^\infty \dfrac{F \hat{O} F}{V^D - E} dr \label{Integral}
\end{align}
where $\hat O$ represents the operator in different channels. Specifically, the kinetic term (see Fig. \ref{fig.3}) corresponds to $\hat O = d^2/dr^2 + V_{PCB}$.

\section{Results and discussion}

In this work, we focus on the even-even $N=32$ and $34$ isotones from Si ($Z=14$) to Ni ($Z=28$) and the calculations are performed within the relativistic Hartree-Fock (RHF) scheme by imposing the spherical symmetry to all the selected nuclei. The RHF mean fields are evaluated by the covariant density functional with Fock terms PKA1 \cite{W.H.Long-2007-PRC}, which takes the one-pion exchange and $\rho$-tensor couplings into account via the Fock terms and interprets successfully the new magicities $N=32$ and $34$ in calcium \cite{J.J.Li-2016-PLB}. Moreover, with the presence of Fock terms, the important ingredient of nuclear force --- tensor force can get involved naturally \cite{L.J.Jiang-2015-PRC, Y.Y.Zong-2018-CPC}.

For the open-shell isotones, the pairing correlations are handled with the BCS approximation, and the pairing force is adopted as the finite-range Gogny D1S \cite{Berger-1984-NPA} to have a natural energy cutoff. It should be noticed that although some isotones like $^{48}$Si are located at the neutron drip line, the BCS pairing with the finite-range pairing force Gogny D1S can still provide reasonable evaluation for the structure properties of the relevant exotic nuclei due to the predicted magicities \cite{J.J.Li-2019-PLB}. In fact, the calculations with Bogoliubov pairing \cite{J.Meng-1996-PRL, Dobaczewski-1984-NPA, J.Meng-1998-NPA, S.G.zhou-2010-PRC, L.L.Li-2012-PRC, X.X.Sun-2018-PLB}, namely by the relativistic Hartree-Fock-Bogoliubov (RHFB) theory \cite{W.H.Long-2010-PRC}, present similar systematics on the PSS restoration along the selected isotonic chains. It is not very surprising to have such consistence. Because of the predicted neutron and proton magic shells in $^{48}$Si \cite{J.J.Li-2019-PLB}, little continuum effect is involved, which indeed plays an essential role in other exotic nuclei, such as for the halo formation \cite{J.Meng-1996-PRL, S.G.zhou-2010-PRC, L.L.Li-2012-PRC, X.X.Sun-2018-PLB}. On the other hand, the localization of the non-local integral terms [see Eq.(3)] does not work anymore for solving directly the RHFB equation, which is instead solved by expanding the quasi-particle wave functions on the Dirac Woods-Saxon basis \cite{S.G.Zhou-2003-PRC, W.H.Long-2010-PRC}. Technically, it is more convenient and straightforward to understand the physics in the PSS restoration under the BCS scheme.

\begin{center}
\ifpdf
\includegraphics[width=0.4\textwidth]{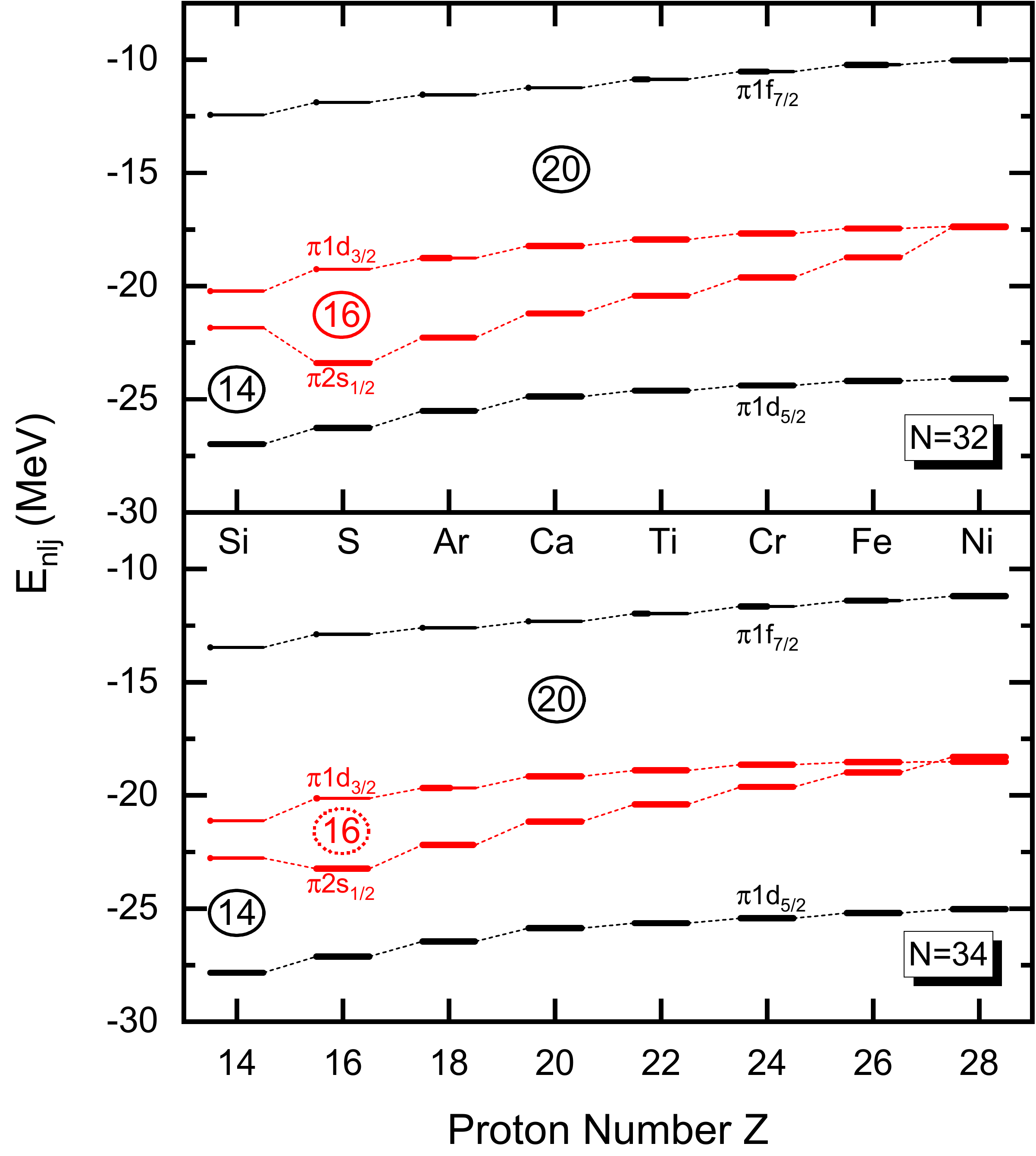}
\else
\includegraphics[width=0.4\textwidth]{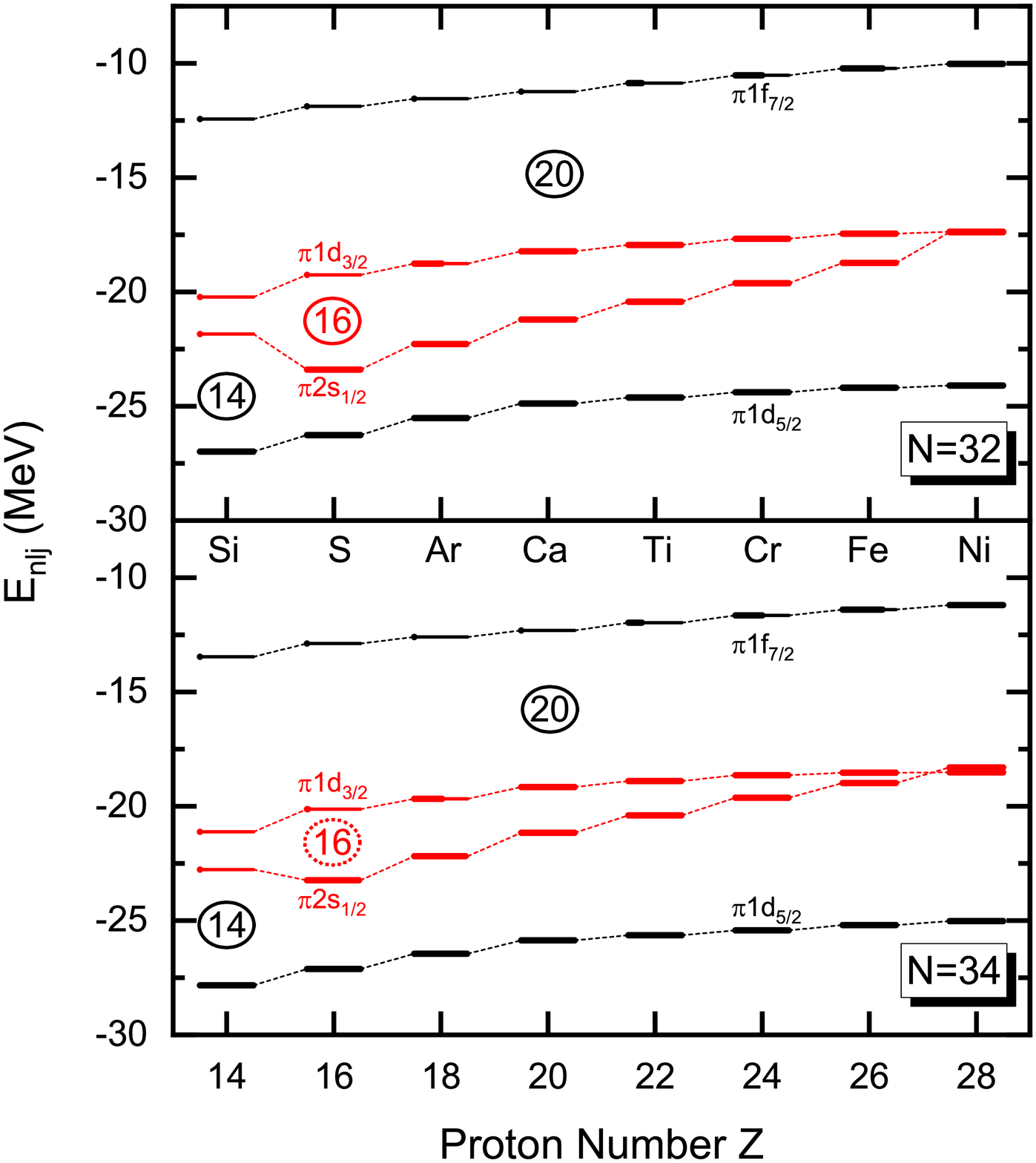}
\fi
\figcaption{ \label{fig.1}Proton single-particle energies of $N=32$ (upper panel) and $N=34$ (lower panel) isotones from Si to Ni, where the thick bars denote the occupation probabilities. The results are extracted from the calculations of RHF with PKA1 \cite{W.H.Long-2007-PRC}, in which the pairing correlations are considered with BCS approximation plus the finite-range pairing force Gogny D1S \cite{Berger-1984-NPA}.}
\end{center}

Fig. \ref{fig.1} shows the proton ($\pi$) single-particle energies along the isotonic chains of $N=32$ and $34$ from silicon to nickel, and the results are obtained by the RHF theory with PKA1 plus finite-range pairing force Gogny D1S. It can be seen from Fig. \ref{fig.1} that the single-particle energies monotonously increase with respect to the proton number $Z$, except the $\pi2s_{1/2}$ orbits which become deeper bound from silicon to sulphur. As a result of the abnormal behavior of $\pi2s_{1/2}$, the sub-shell $Z=14$, new magic shell confirmed experimentally in $^{34}$Si \cite{Mutschler-2017-nature-Si34} and predicted theoretically in $^{48}$Si \cite{J.J.Li-2019-PLB}, are quenched distinctly, leading to the emergence of a remarkable sub-shell $Z=16$ in $^{48}$S ($N=32$) and less soundable one in $^{50}$S ($N=34$) deduced from the distinct violation of the PSS for the doublet $(\pi2s_{1/2}, \pi1d_{3/2})$. Afterwards from sulphur to nickel, the PSO splitting between $\pi2s_{1/2}$ and $\pi1d_{3/2}$ monotonously decreases with the proton number, eventually presenting appropriate restoration of the PSS at $^{60, 62}$Ni.

As seen from Fig. \ref{fig.1}, the proton valence orbits for the selected isotones are the pseudo-spin partners $(\pi2s_{1/2}, \pi1d_{3/2})$ and $\pi1f_{7/2}$, which contain the $j = l-1/2$ and $j=l+1/2$ orbits. For the neutron valence orbits, mainly the spin partners $\nu2p$ and the state $\nu1f_{5/2}$, the single-particle configuration will be slightly changed because of the pairing effects in some open-shell isotones. Thus, it is naturally expected that the tensor force might play a role in the PSS restoration, due to its characteristic nature --- the spin dependence \cite{Otsuka-2010-PRL, W.H.Long-2008-EPL, W.H.Long-2010-PRC, J.J.Li-2016-PLB, L.J.Jiang-2015-PRC}. Starting from the Dirac equation (\ref{Dir}), Fig. \ref{fig.2} shows the evolution of pseudo-spin orbital splittings of the doublets $(\pi2s_{1/2}, \pi1d_{3/2})$, denoted as $\Delta E_{\pi1p'} = E_{\pi1d_{3/2}} - E_{\pi2s_{1/2}}$, along the $N=32$ and $34$ isotonic chains. It can be clearly seen that distinct PSS violation is found in both $N=32$ and $34$ isotones of sulphur, and towards nickel the PSS becomes well restored, even leading to the inversion of $sd$ orbits at nickel. With the relativistic formalism proposed in Ref. \cite{L.J.Jiang-2015-PRC} and implemented in Ref. \cite{Y.Y.Zong-2018-CPC}, the contributions from the tensor force components, which are naturally introduced by the Fock terms, are extracted and shown with open symbols in Fig. \ref{fig.2}. Being consistent with the nature of the tensor force and the systematics of the proton configurations shown in Fig. \ref{fig.1}, the tensor contributions to the pseudo-spin orbit splitting $\Delta E_{\pi1p'}$ are weakened when approaching the spin-saturated proton $Z=20$, namely $^{52,54}$Ca, and afterwards such effects tend to be enhanced since protons become spin-unsaturated. Even though, the tensor force cannot account for the overall systematics of the PSS restoration along both isotonic chains, which only partly interpret the evolution of $\Delta E_{\pi1p'}$ from calcium to nickel.

\begin{center}
\ifpdf
\includegraphics[width=0.4\textwidth]{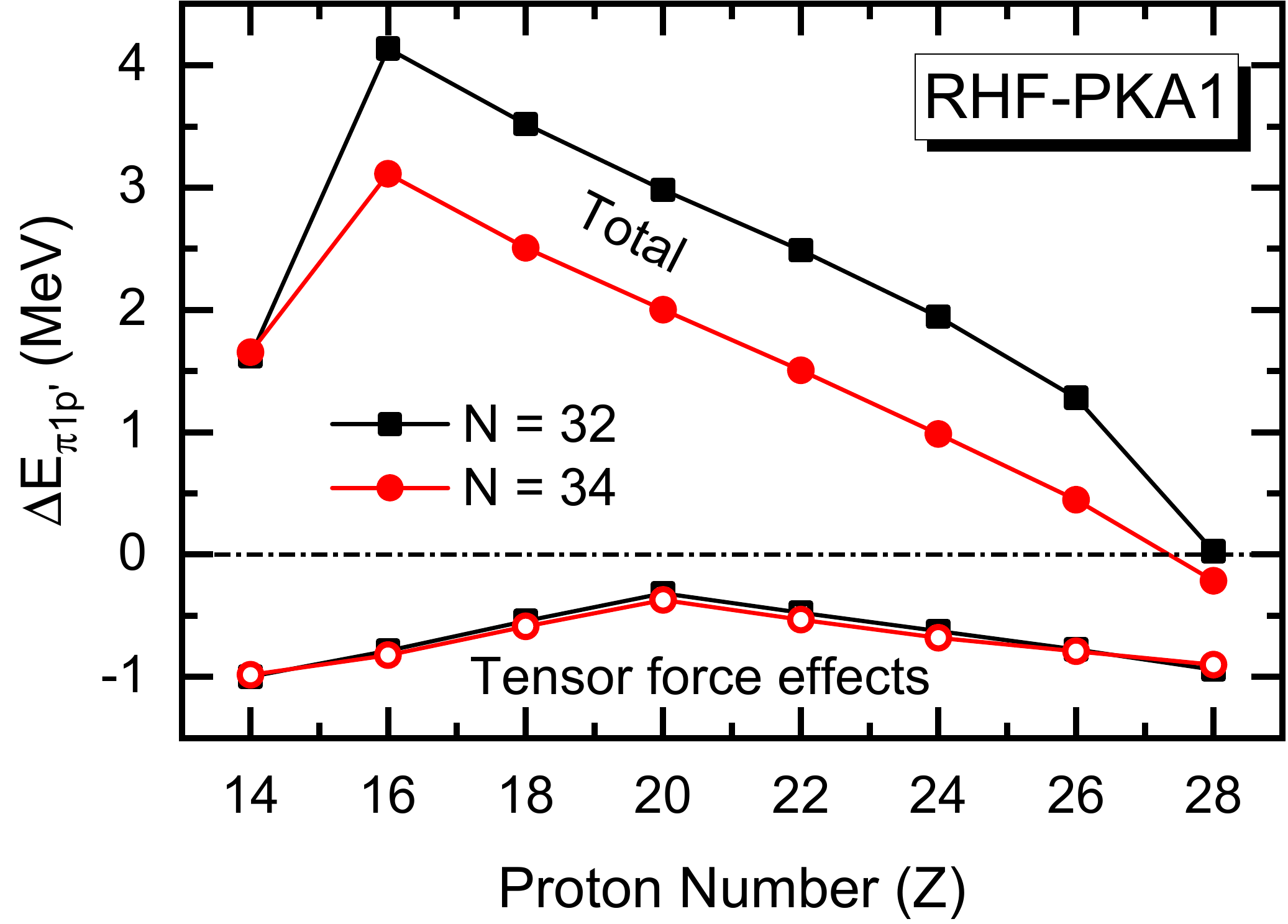}
\else
\includegraphics[width=0.4\textwidth]{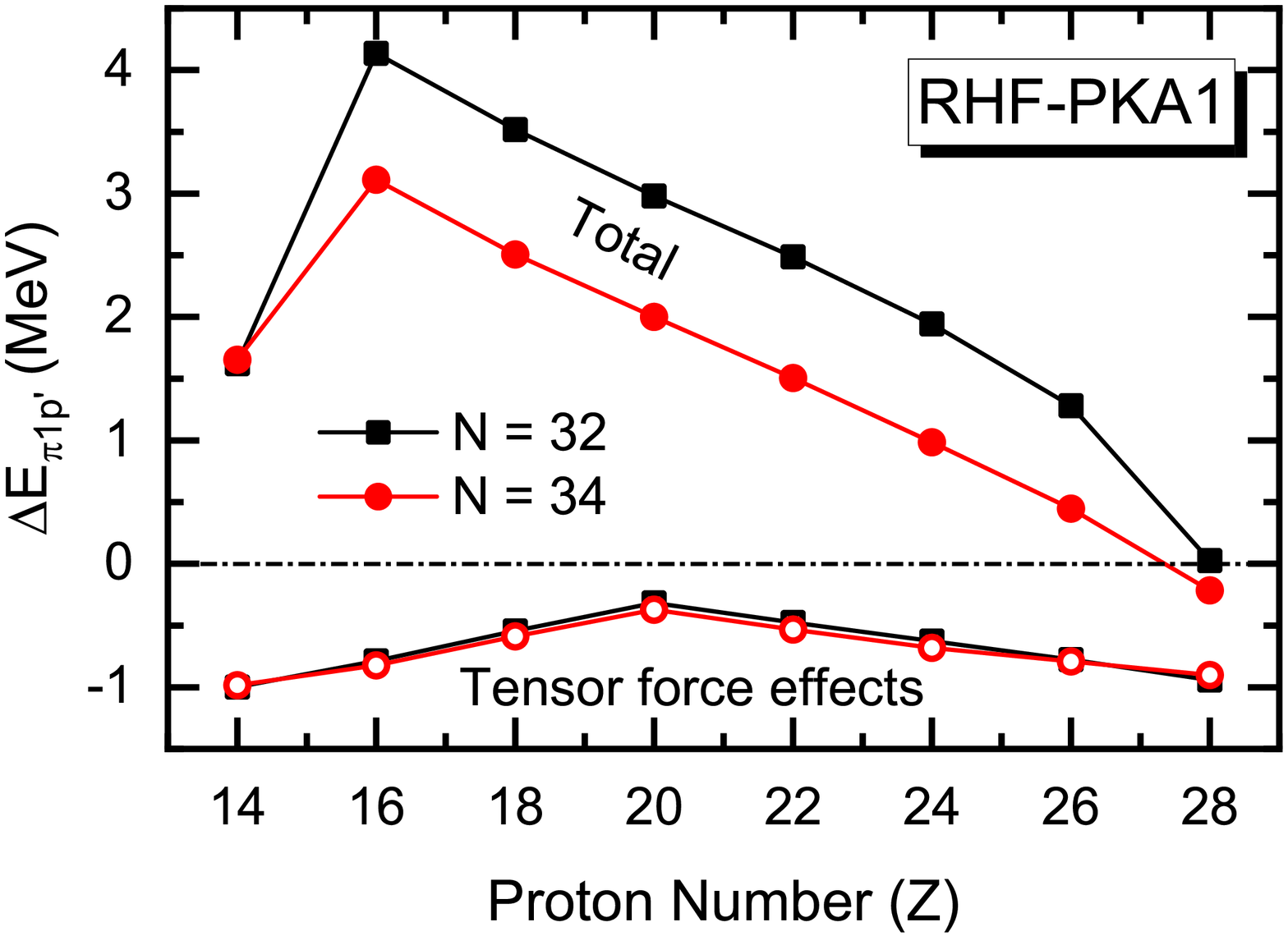}
\fi
\figcaption{ \label{fig.2}Evolution of the pseudo-spin orbit splitting $\Delta E_{\pi 1p'} = E_{\pi2s_{1/2}}- E_{\pi1d_{3/2}}$ along the isotonic chains of $N=32$ and $N=34$, where the open symbols represent the contributions from the tensor forces. The results are extracted from the calculations of RHF with PKA1 plus finite-range pairing force Gogny D1S. }
\end{center}

In order to better understand the overall evolution of $\Delta E_{\pi1p'}$ along both isotonic chains in Fig. \ref{fig.2}, particularly for the distinct violation of PSS at sulphur, the radial Dirac equation is reduced into a Schr\"{o}dinger-like one for the lower component of Dirac spinor as people did in Refs. \cite{J.Meng-1998-PRC, W.H.Long-2006-PLB-PSS}, namely Eq. (\ref{Sch-L}). It can give quantitative evaluation on the contributions to $\Delta  E_{\pi1p'}$ from various channels. In practice, the contributions are detailed as the kinetic, Hartree and Fock terms, and the obtained results are given in Fig. \ref{fig.3}, showing similar systematics for $N=32$ (left panel) and $34$ (right panel) isotones. From Fig. \ref{fig.3}, it is found that the overall evolution of $\Delta E_{\pi1p'}$, including the distinct violation at sulphur, are largely determined by the Hartree term, which are partly canceled by the kinetic and Fock terms. In fact, it is not quite surprising to have such results since the tensor force, that contributes via the Fock terms, does not play the role as one expected. As well known, the PSS as a relativistic symmetry originates from the large cancellation between the strong scalar and vector fields, namely $V+S = 0$ \cite{Ginocchio-1997-PRL} or $d(V+S)/dr= 0$ \cite{J.Meng-1998-PRC}. Although with the presence of Fock terms the situation becomes more complicated in the RHF theory \cite{W.H.Long-2006-PLB-PSS}, the Lorentz covariance still remains and the opposite behaviors between Hartree and Fock terms in Fig. \ref{fig.3} are simply due to the opposite sign from the anti-commutation relation of Fermions.

From Fig. \ref{fig.3} the overall systematics of $\Delta E_{\pi1p'}$ has been well interpreted. While it still remains the puzzle why the distinct PSS violation appears from silicon to sulphur for both $N=32$ and $34$ isotones. As a further exploration, Fig. \ref{fig.4} shows the proton density profiles of silicon, sulphur, calcium, chromium and nickel, as well as the mean field potentials $\Delta ^D  = \Sigma_S+\Sigma_0$  whose radial derivatives~ together~ with~ $\kappa$-quantities~ dominate~ the

\begin{center}
\ifpdf
\includegraphics[width=0.45\textwidth]{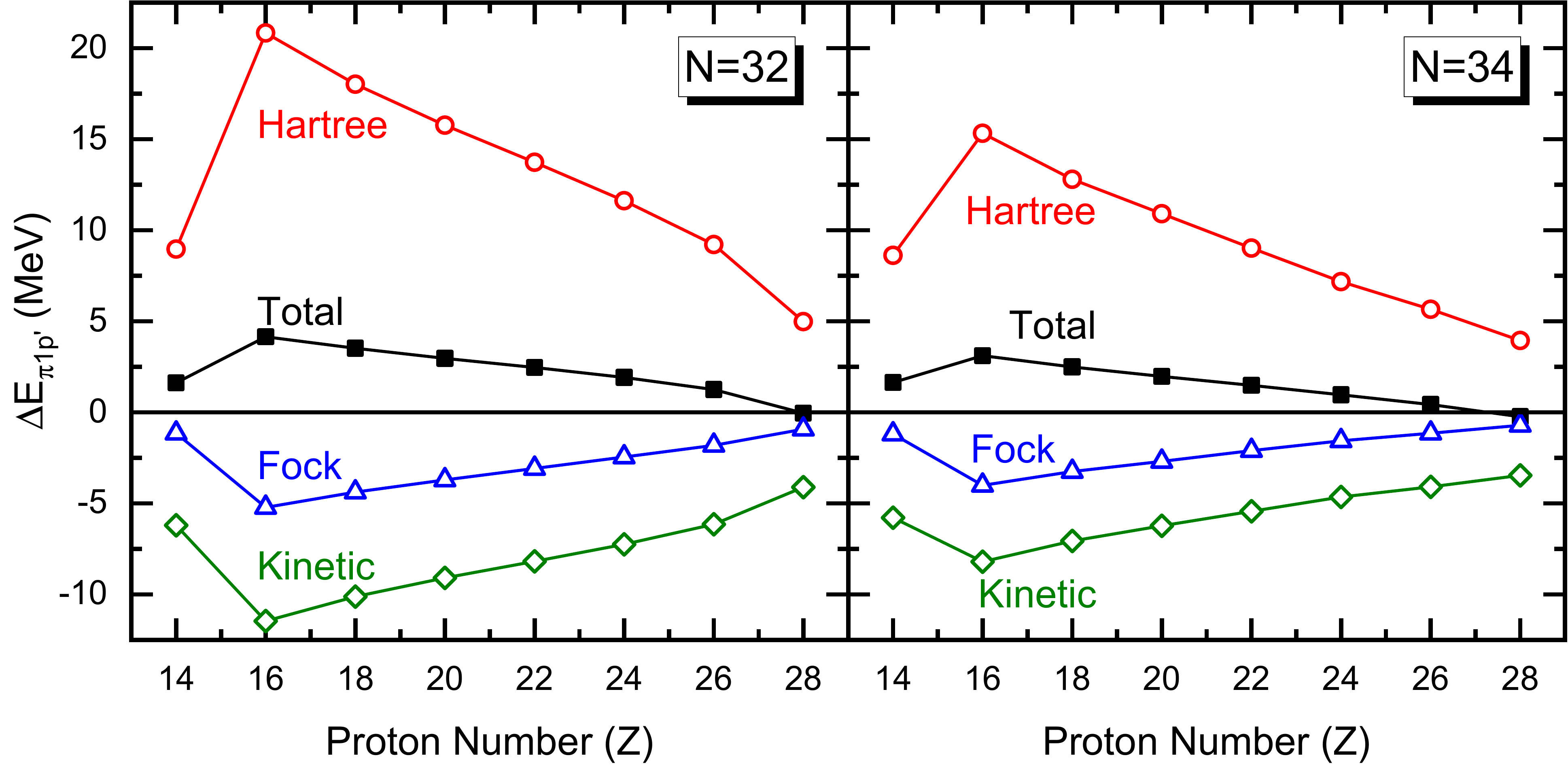}
\else
\includegraphics[width=0.45\textwidth]{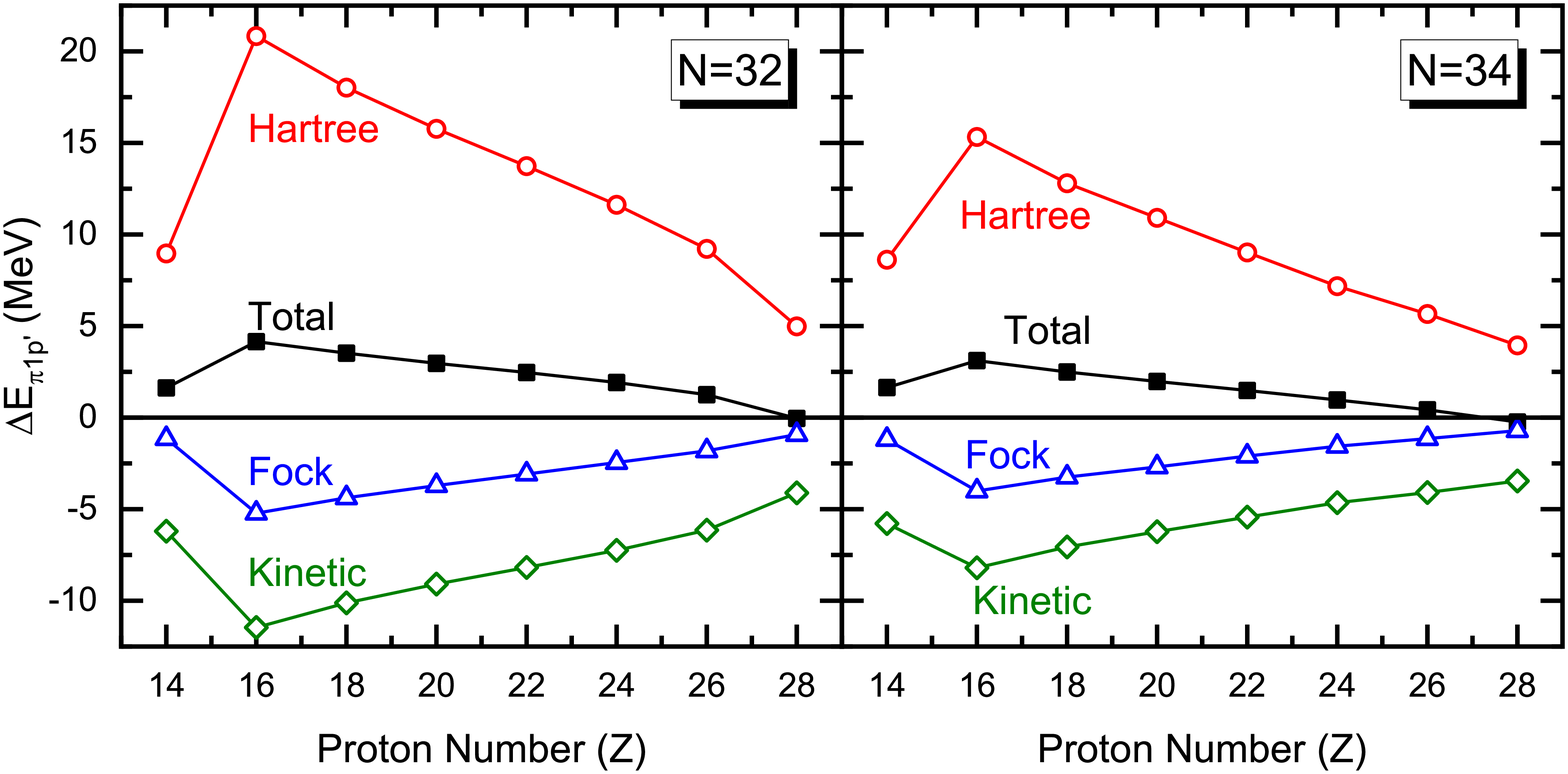}
\fi
\figcaption{\label{fig.3}Contributions to the pseudo-spin orbital splittings $\Delta E_{\pi1p'}$ along the isotonic chains of $N=32$ (left panel) and $N=34$ (right panel), including the total and the ones from the Kinetic, Hartree and Fock terms. The results are extracted from the calculations of RHF with PKA1 plus finite-range pairing force Gogny D1S.}
\end{center}

\noindent strength of pseudo-spin orbital couplings. It can be seen that the proton density profiles of $^{46,48}$Si are central-depressed distinctly, because the $\pi2s_{1/2}$ orbits are empty as one can see from Fig. \ref{fig.1}. Such semi-bubble structure, as a quantum effect, can however be reduced by the correlations beyond the mean field. For instance, if the $s$-state is close enough to the last occupied one, the pairing correlations \cite{J.J.Li-2016-PRC, Khan-2008-NPA, Grasso-2009-PRC, Nakada-2013-PRC}, as well as the multi-reference framework beyond mean field \cite{J.M.Yao-2012-PRC, J.M.Yao-2013-PLB}, may lead to substantially occupied $s$-state that can wash out the central depletion. Even though, the sizes of the energy gaps $Z=14$ in $^{46,48}$Si (see Fig. \ref{fig.1}) are large enough to assess the prediction of a bubble-like structure. With the $\pi2s_{1/2}$ orbits are fully occupied in $^{48, 50}$S, the density profiles becomes distinctly central-bumped. Consistently, the mean field potentials $\Delta^D$ in $^{48, 50}$S are deepened remarkably. Although the Coulomb repulsion is considerably enhanced from silicon to sulphur, the deepened mean field potential leads to deeper bound $\pi2s_{1/2}$ in sulphur, as seen in Fig. \ref{fig.1}, because the $s$ orbits, without the blocking of centrifugal repulsion, can strongly couple with the interior region of $\Delta^D$. On the contrary, the $\pi1d_{3/2}$ orbits in $^{48, 50}$S, with strong centrifugal repulsion, are less bound than in $^{46,48}$Si. Thus, the pseudo-spin orbital splitting $\Delta E_{\pi1p'}$ is distinctly enlarged from silicon to sulphur.

In fact, referring to the conservation condition of the PSS, i.e., $d[V(r)+S(r)]/dr = 0$, one can also well understand the evolution of $\Delta E_{\pi1p'}$ from silicon to sulphur. As shown in Fig.\ref{fig.4} (c), the mean field potential $\Delta ^D$ in $^{46}$Si is consistently central-bumped with the proton semi-bubble structure [see plot (a)], which provides an cancellation to the derivative $d \Delta^D /dr$ at the edge of the potential well. On the contrary, with two more protons occupying the $\pi2s_{1/2}$ orbit, the proton density in $^{48}$S becomes central-bumped and consistently the potential $\Delta^D$ is central-depressed, which presents distinct enhancement to the derivative $d \Delta^D / dr$. Thus, the PSO potential, which is proportional to the derivative $d\Delta^D/dr$, is enhanced, leading to distinctly enlarged PSO splitting $\Delta E_{\pi1p'}$ from $^{46}$Si to $^{48}$S. To less extent, similar situation can be found in the N=34 isotones $^{48}$Si and $^{50}$S, see Figs.\ref{fig.4} (b) and (d). After sulphur, with the valence protons gradually occupy the $\pi1d_{3/2}$ and $1f_{7/2}$ orbits, the density profiles become less and less central-bumped and consistently the mean field potential $\Delta^D$ become more and more central-flat, leading to eventually well-restored PSS at nickel. Additionally, within the RMF theory more realistic condition, i.e., $V_{PCB}>>V_{PSO}$, was proposed to guarantee the approximate PSS conservation in nuclei \cite{J.Meng-1998-PRC}. While within the RHF theory, the situation is largely changed and such condition is not fulfilled any more due to the complicated non-local mean field induced by the Fock terms \cite{W.H.Long-2006-PLB-PSS}, which has been also tested for the selected isotones.

\begin{center}
\ifpdf
\includegraphics[width=0.45\textwidth]{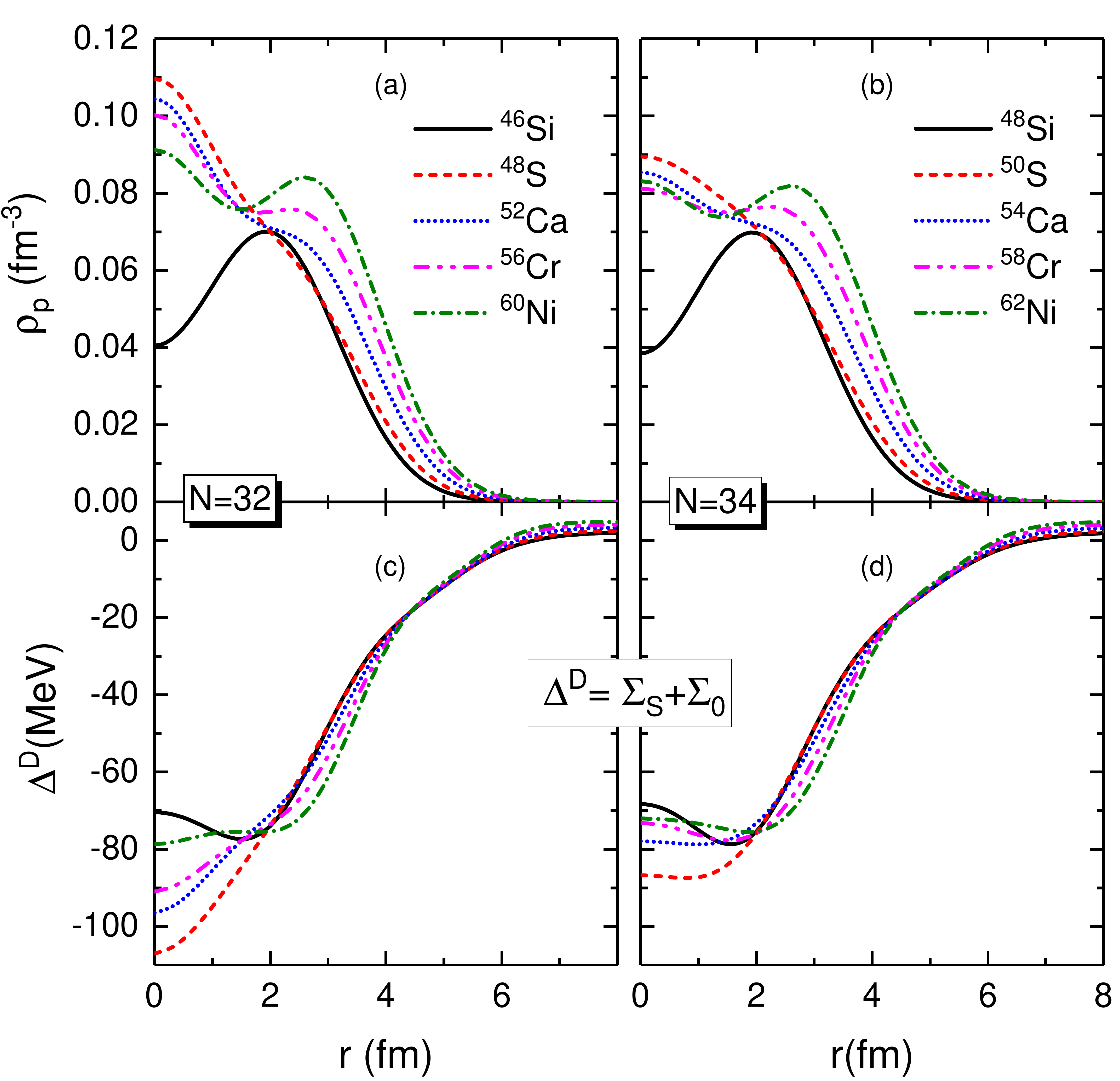}
\else
\includegraphics[width=0.45\textwidth]{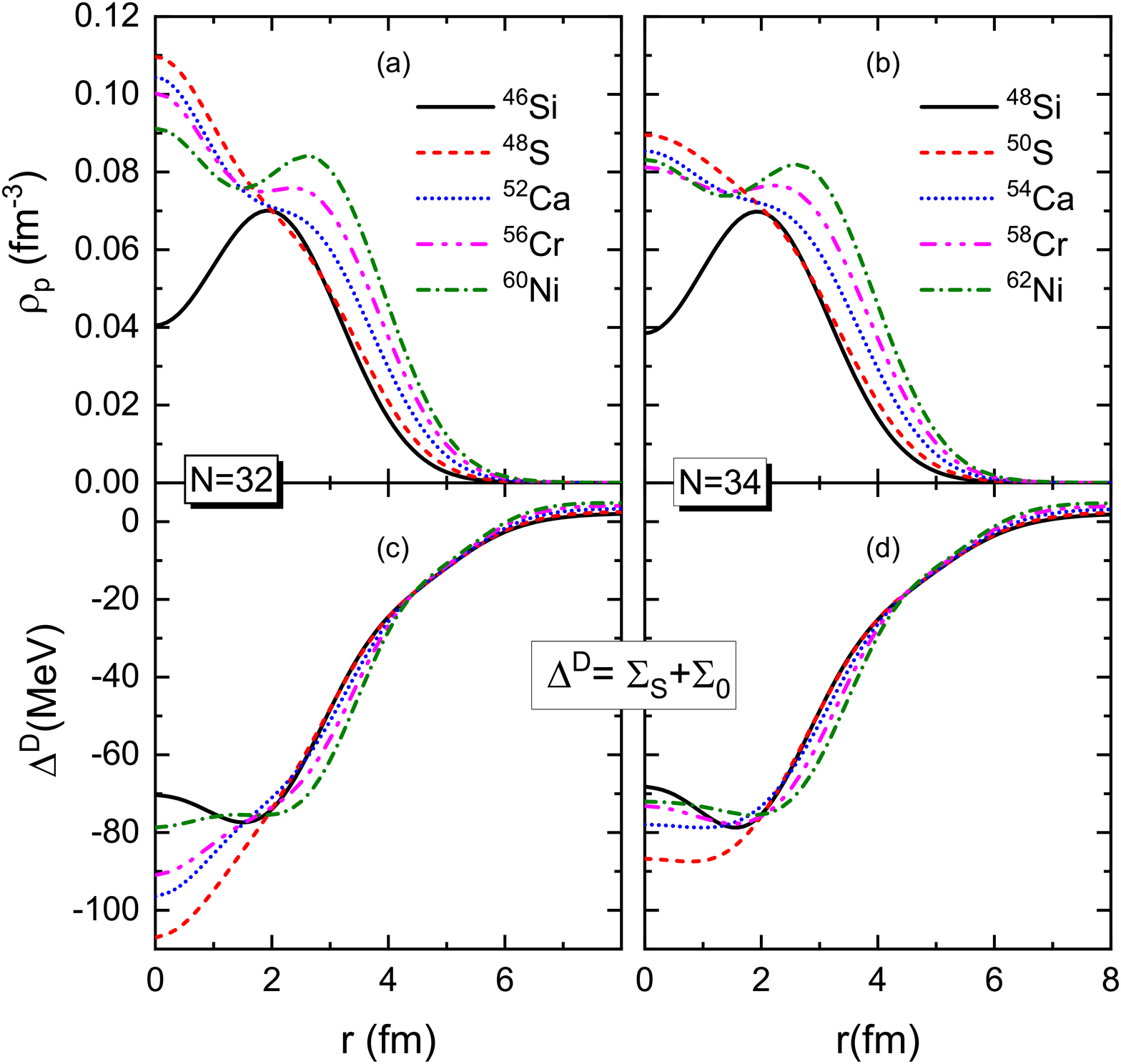}
\fi
\figcaption{\label{fig.4} Proton densities [plots (a,b)] and the mean field potential $\Delta^D = \Sigma_S + \Sigma_0$ [plots (c, d)] of the selected $N=32$ (left panels) and $N=34$ (right panels) isotones. The results are extracted from the calculations of RHF with PKA1 plus finite-range pairing force Gogny D1S.}
\end{center}

\section{Summary}
In summary, in this work the restoration of the pseudo-spin symmetry (PSS) along the $N=32$ and $34$ isotonic chains, as well as the driving mechanism, is studied by applying the relativistic Hartree-Fock calculations with PKA1 and BCS approximation for the pairing correlations with finite range pairing force Gogny D1S. In both $N=32$ and $34$ isotones, distinct violation of PSS appears from silicon to sulphur for the proton pseudo-spin doublets $(\pi2s_{1/2}, \pi1d_{3/2})$, which opens the sub-shell $Z=16$ in $^{48}$S, and from sulphur to nickel the PSS becomes better and better restored, leading to the inversion of $sd$ orbits at nickel. It is found that the effects of tensor force introduced naturally via the Fock terms can only partly interpret the restoration of the PSS from calcium to nickel, while fail for the overall systematics along both isotonic chains. By reducing radial Dirac equation into the Schr\"{o}dinger-like one, further investigation illustrates that the overall systematics of the restoration of the PSS can be well explained by the contributions of Hartree terms. Moreover, the analysis of the proton density profiles and relevant mean field potentials indicates that the PSS violation at sulphur can be attributed to dramatic changes of the bubble-like density profiles in silicon to the central-bumped ones in sulphur, and afterwards the PSS restoration can be self-consistently interpreted by more and more central-flat density profile from sulphur to nickel, as well as the consistent changes of the mean field potentials.

\end{multicols}
\vspace{10mm}
\begin{multicols}{2}

\end{multicols}

\end{document}